\documentclass[12pt,preprint]{aastex}

\shorttitle{Constraints on the VHE Emission from BL Lacs}
\shortauthors{Horan D., et al.}

\begin{document}

\title{Constraints on the Very High Energy Emission from BL Lacertae
Objects}

\author{D. Horan,\altaffilmark{1}
H. M. Badran, \altaffilmark{2}
I. H. Bond,\altaffilmark{3}
P. J. Boyle,\altaffilmark{4}
S. M. Bradbury,\altaffilmark{3}
J. H. Buckley,\altaffilmark{5}
D. A. Carter-Lewis,\altaffilmark{6}
M. Catanese,\altaffilmark{1}
O. Celik,\altaffilmark{7}
W. Cui,\altaffilmark{8}
M. Daniel,\altaffilmark{8}
M. D'Vali,\altaffilmark{3}
I. de la Calle Perez,\altaffilmark{3}
C. Duke,\altaffilmark{9}
A. Falcone,\altaffilmark{8}
D. J. Fegan,\altaffilmark{10}
S. J. Fegan,\altaffilmark{1}
J. P. Finley,\altaffilmark{8}
L. F. Fortson,\altaffilmark{4}
J. A. Gaidos,\altaffilmark{8}
S. Gammell,\altaffilmark{10}
K. Gibbs,\altaffilmark{1}
G. H. Gillanders,\altaffilmark{11}
J. Grube,\altaffilmark{3}
J. Hall,\altaffilmark{12}
T. A. Hall,\altaffilmark{13}
D. Hanna,\altaffilmark{14}
A. M. Hillas,\altaffilmark{3}
J. Holder,\altaffilmark{3}
A. Jarvis,\altaffilmark{7}
M. Jordan,\altaffilmark{5}
G. E. Kenny,\altaffilmark{11}
M. Kertzman,\altaffilmark{15}
D. Kieda,\altaffilmark{12}
J. Kildea,\altaffilmark{15}
J. Knapp,\altaffilmark{3}
K. Kosack,\altaffilmark{5}
H. Krawczynski,\altaffilmark{5}
F. Krennrich,\altaffilmark{6}
M. J. Lang,\altaffilmark{11}
S. Le Bohec,\altaffilmark{6}
E. Linton,\altaffilmark{4}
J. Lloyd-Evans,\altaffilmark{3}
A. Milovanovic,\altaffilmark{3}
P. Moriarty,\altaffilmark{16}
D. Muller,\altaffilmark{4}
T. Nagai,\altaffilmark{12}
S. Nolan,\altaffilmark{8}
R. A. Ong,\altaffilmark{7}
R. Pallassini,\altaffilmark{3}
D. Petry,\altaffilmark{17}
B. Power-Mooney,\altaffilmark{10}
J. Quinn,\altaffilmark{10}
M. Quinn,\altaffilmark{16}
K. Ragan,\altaffilmark{14}
P. Rebillot,\altaffilmark{5}
P. T. Reynolds,\altaffilmark{18}
H. J. Rose,\altaffilmark{3}
M. Schroedter,\altaffilmark{1}
G. H. Sembroski,\altaffilmark{8}
S. P. Swordy,\altaffilmark{4}
A. Syson,\altaffilmark{3}
V. V. Vassiliev,\altaffilmark{12}
S. P. Wakely,\altaffilmark{4}
G. Walker,\altaffilmark{12}
T. C. Weekes,\altaffilmark{1}
J. Zweerink,\altaffilmark{7}}

\email{dhoran@cfa.harvard.edu} 

\altaffiltext{1}{Fred Lawrence Whipple Observatory, Harvard-Smithsonian CfA, P.O. Box 97, Amado, AZ 85645-0097} 
\altaffiltext{2}{Physics Department, Tanta University, Tanta, Egypt}
\altaffiltext{3}{Department of Physics, University of Leeds, Leeds, LS2 9JT, Yorkshire, England, UK}
\altaffiltext{4}{Enrico Fermi Institute, University of Chicago, Chicago, IL 60637, USA}
\altaffiltext{5}{Department of Physics, Washington University, St. Louis, MO 63130, USA}
\altaffiltext{6}{Department of Physics and Astronomy, Iowa State University, Ames, IA 50011-3160, USA}
\altaffiltext{7}{Department of Physics, University of California, Los Angeles, CA 90095-1562, USA}
\altaffiltext{8}{Department of Physics, Purdue University, West Lafayette, IN 47907, USA}
\altaffiltext{9}{Department of Physics, Grinnell College, Grinnell, IA 50112-1690, USA}
\altaffiltext{10}{Experimental Physics Department, National University of Ireland, Belfield, Dublin 4, Ireland}
\altaffiltext{11}{Department of Physics, National University of Ireland, Galway, Ireland}
\altaffiltext{12}{High Energy Astrophysics Institute, University of Utah, Salt Lake City, UT 84112, USA}
\altaffiltext{13}{Department of Physics and Astronomy, University of Arkansas at Little Rock, Little Rock, AR 72204-1099, USA}
\altaffiltext{14}{Physics Department, McGill University, Montre$\acute{a}$l, QC\,H3A\,2T8, Canada}
\altaffiltext{15}{Department of Physics and Astronomy, DePauw University, Greencastle, IN 46135-0037, USA}
\altaffiltext{16}{School of Science, Galway-Mayo Institute of Technology, Galway, Ireland}
\altaffiltext{17}{University of Maryland, Baltimore County and NASA/GSFC, USA}
\altaffiltext{18}{Department of Applied Physics and Instrumentation, Cork Institute of Technology, Cork, Ireland}

\begin{abstract} 

We present results from observations of 29 BL Lacertae objects, taken
with the Whipple Observatory 10\,m Gamma-Ray Telescope between 1995
and 2000.  The observed objects are mostly at low redshift ($z < 0.2$)
but observations of objects of $z$ up to 0.444 are also reported. Five
of the objects are EGRET sources and two are unconfirmed TeV
sources. Three of the confirmed sources of extragalactic TeV gamma
rays were originally observed as part of this survey and have been
reported elsewhere. No significant excesses are detected from any of
the other objects observed, on time scales of days, months or
years. We report 99.9\% confidence level flux upper limits for the
objects for each observing season. The flux upper limits are typically
20\% of the Crab flux although, for some sources, limits as sensitive
as 6\% of the Crab flux were derived. The results are consistent with
the synchrotron-self-Compton (SSC) model predictions considered in
this work.

\end{abstract}


\section{Introduction}
\label{intro}

BL Lacertae objects (BL Lacs) are members of the blazar class of
active galactic nuclei (AGN). Like all blazars, they exhibit rapid,
large amplitude variability at all wavelengths, high optical and radio
polarization, and in some cases, apparent superluminal motion and/or
gamma-ray emission. All of these observational properties lead to the
widely held belief that blazars are AGN with jets oriented nearly
along our line of sight \citep{Urry95}. The broadband spectral energy
distribution (SED) of blazars, when plotted as $\nu$F$_\nu$ versus
frequency, shows a double peaked shape, with a smooth extension from
radio to between IR and X-ray frequencies (depending on the specific
blazar type), followed by a distribution that typically starts in the
X-ray band and can peak in the gamma-ray band, at energies as high as
several hundred GeV. EGRET has detected more than 65 blazars
\citep{Hartman99}, 14 of which have been identified as BL Lacs
\citep{Dermer00}.

The low energy part of the blazar SED is believed to be incoherent
synchrotron radiation from a relativistic electron-positron plasma in
the blazar jet. The origin of the high energy emission is still a
matter of considerable debate (e.g., \citealp{Buckley98};
\citealp{Mannheim98}).  Leptonic models are the most popular models
used to explain the observed gamma-ray emission from blazars. In
synchrotron-self-Compton (SSC) models, the gamma rays are produced
through inverse Compton scattering of low energy photons by the same
electrons that produce the synchrotron emission at lower energies
\citep{Konigl81,Maraschi92,Dermer92,Sikora94,Bloom96,Sikora01}.
In external Compton models, the dominant source of seed photons for
upscattering in the inverse Compton process are ambient photons from
the central accretion flow, the accretion disk, the broad line region,
the torus, the local infra-red background, or some combination of
these \citep{Sikora94, Dermer93, Blandford95, Ghisellini96,
Wagner95}. Another set of models proposes that the gamma rays are
the result of ultra-high energy (E $\gtrsim 10^{19}$ eV) protons
producing TeV gamma rays as proton synchrotron radiation
\citep{Aharonian00}, in proton induced electromagnetic cascades
\citep{Mannheim98}, or by a combination of both processes
\citep{Mucke03}.

Among blazars, BL Lacs are believed to be the best candidates for TeV
emission. They have weak or absent optical emission lines, indicating
that they may have less TeV-absorbing material near the emission
region \citep{Dermer94}. Although they show relatively low
luminosities compared to Flat Spectrum Radio Quasars (FSRQs) and
Optically Violent Variables (OVVs), their SEDs peak at higher energies
\citep{Fossati98}. In blazar unification models in which electrons are
assumed to be the progenitors of the gamma rays
(e.g. \citealp{Ghisellini98}), the lower luminosity of the BL Lacs,
relative to the FSRQs implies that in BL Lacs, electrons are cooled
less efficiently. This means that the electrons reach higher energies
and that the emitted gamma rays will therefore also be of higher
energy.

Consistent with these expectations, BL Lacs are the only type of
blazars detected at very high energies (VHE, E $\gtrsim$
250\,GeV). Currently, six BL Lacs are confirmed sources of VHE gamma
rays, Mrk\,421 \citep{Punch92}, Mrk\,501 \citep{Quinn96}, H\,1426+428
\citep{Horan02} and 1ES\,1959+650 \citep{Nishiyama00}; the original
detection of the nearby BL Lac, 1ES\,2344+514 \citep{Catanese98} was
confirmed by the HEGRA group \citep{Tluczykont03} while that of the
Southern Hemisphere BL Lac, PKS\,2155-304 \citep{Chadwick99} was
confirmed recently by H.E.S.S. \citep{Djannati03}. Unconfirmed
detections have been reported for the BL Lac objects 3C\,66A
\citep{Neshpor98} and BL Lacertae \citep{Neshpor01}. The TeV blazars,
1ES\,2344+514, H\,1426+428 and 1ES\,1959+650 were originally observed
at Whipple as part of the BL Lac survey described here. Since they
were detected during these observations, the results were described in
detail elsewhere \citep{Catanese98,Horan02,Holder03} and are
summarized here.

\citet{Padovani95a} introduced the terminology ``Low-frequency peak BL
Lacs'' (LBLs) to describe those BL Lacs in which the lower energy SED
peak occurs in the radio band and ``High-frequency peaked BL Lacs''
(HBLs) for those whose lower energy peak occurs in the X-ray
band. Recently, deeper BL Lacs surveys (e.g. \citealp{Perlman98};
\citealp{Laurent99}; \citealp{Caccianiga99}) have revealed evidence
for the existence of BL Lacs with properties intermediate to those in
the LBL and HBL classes. Indeed, W\,Comae has recently been classified
as an intermediate BL Lac \citep{Tagliaferri00}.  This suggests that,
rather than being separate subclasses, LBLs and HBLs represent the
edges of a sequence of progressively different BL Lacs. The term
``extreme blazars'' was introduced by \citet{Ghisellini99} to describe
those BL Lacs whose first peak extends into the hard X-ray band. Such
objects, which lie at the end of the ``blazar sequence'' proposed by
\citet{Fossati97}, are good candidates for TeV emission since the
second peak in their SEDs also lies at higher energies meaning that
they can be powerful at TeV energies.

With the exception of 3C\,66A and BL Lacertae, all of the claimed and
confirmed TeV gamma-ray emitting BL Lacs are HBLs. H\,1426+428,
Mrk\,501, 1ES\,2344+514 and, to a lesser extent, 1ES\,1959+650, all
have very hard X-ray energy spectra and fall into the class of extreme
blazars. All of the TeV blazars but these four were listed as
detections in the Third EGRET catalog \citep{Hartman99}. The second
peak in the SED of these and other extreme objects can only be studied
in detail at TeV energies as it lies above 100 GeV, where EGRET and
GLAST are less sensitive. Ground-based gamma-ray telescopes therefore
offer a unique opportunity to study this class of higher-peaked
blazars. Indeed, the majority of blazars detected by EGRET were FSRQs
whose second SED peak falls in the MeV to GeV band.

In order to improve our understanding of the gamma-ray emission from
BL Lacs, we need to detect more of these objects at very high
energies. Several groups have published upper limits on VHE emission
from BL Lacs in the last several years
\citep{Roberts98,Roberts99,Chadwick99,Aharonian00}, including the
VERITAS collaboration \citep{Kerrick95}, but these efforts have been
on smaller groups of objects, or not directed specifically at BL
Lacs. In this paper we present the results of our BL Lac observing
program from January 1995 to July 2000. The database comprises
observations of 29 objects totaling 143 hours. We present the results
of searches for emission spanning time-scales of 30 minutes to six
years. No statistically significant excess emission above the
background is found and we discuss the implications of the
non-detections.

These results supersede preliminary analyses presented in conference
proceedings \citep{Catanese97a, Horan00, Horan03}. Five of the
objects presented here were later observed more intensively as part of
a HBL observation program carried out by
\citet{delaCalle03}. In that survey, eight HBLs, selected from a list
of TeV candidate objects derived by \citet{Costamante02}, were
subjected to intensive Whipple observations during 2001 and
2002. These objects were all predicted to be VHE emitters based upon
the location of their synchrotron peak and on their high density seed
photons for the inverse Compton process. No evidence for TeV emission
from these objects was found during this intensive monitoring campaign
\citep{delaCalle03}.

\section{Source List}

\begin{figure}
\epsscale{1}
\plotone{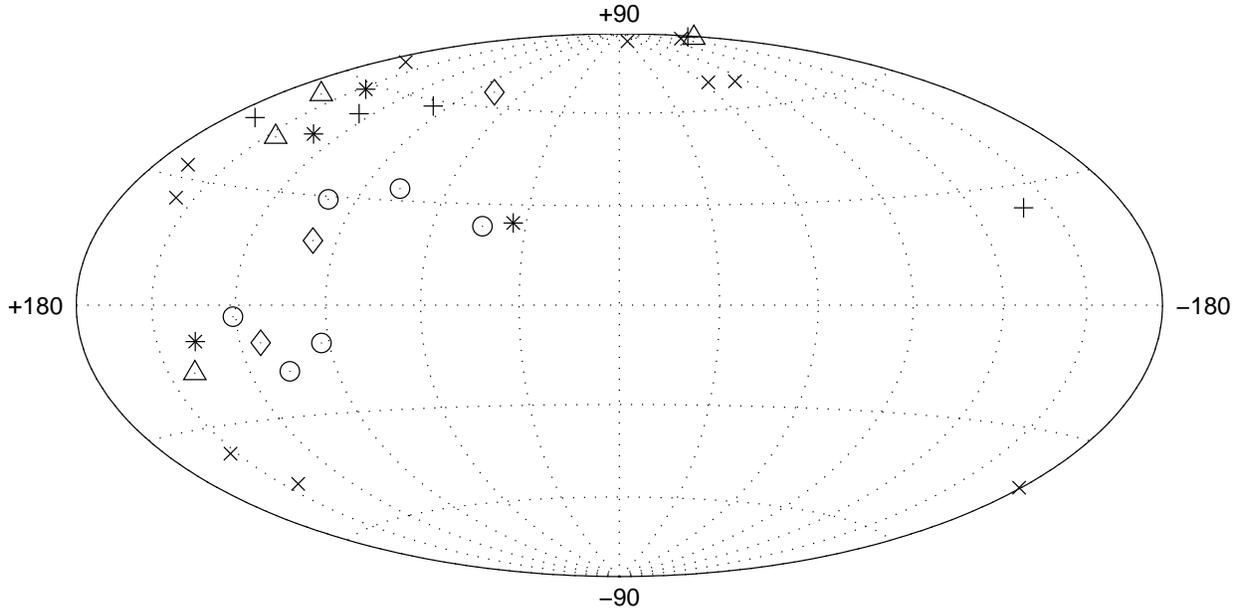}
\caption{\label{skymap}
The locations of the 32 BL Lacs originally observed as part of this
survey are plotted in Galactic co-ordinates. The three objects which
were subsequently detected (1ES\,2344, H\,1426+428 and 1ES\,1959+650)
are labeled with diamonds while the 29 objects whose upper limits are
reported here are labeled according to their redshift. The stars mark
the four objects lying at z $<$ 0.05; the circles mark six objects
lying between z of 0.05 - 0.1; the crosses mark the ten objects lying
between z of 0.1 and 0.15; the plus symbols mark the five objects
lying between z of 0.15 and 0.20; the triangles mark the four objects
which lie at a z $>$ 0.20.}

\end{figure}

Table~\ref{object_sum} lists the objects whose observations are
reported here. In the table, we provide the object name (or names),
its equatorial coordinates, its redshift and classification. The
locations of these BL Lacs, along with those of 1ES\,2344+514,
H\,1426+428 and 1ES\,1959+650, are plotted in Galactic co-ordinates in
Figure~\ref{skymap}.

The majority of the objects in Table~\ref{object_sum} were selected
for observations as part of three BL Lac campaigns. The first was a
survey of all known (circa 1995) BL Lac objects with redshift of
$\lesssim 0.1$. The goal was to investigate the intrinsic
characteristics of BL Lac objects that led to the production of TeV
gamma rays. We selected low redshift objects to minimize the effect of
the attenuation of the VHE gamma-ray signal by pair production with
extragalactic background light \citep{Stecker99,Primack99,Vassiliev00}
so that the intrinsic features could be compared.  The second campaign
was to search for TeV emission from HBLs in the redshift range from
0.1 to 0.2. We believed that because the HBLs have SEDs similar to the
confirmed TeV emitters that they would be stronger TeV candidates than
the LBLs. Because of the potential attenuation by the EBL and the
larger pool of objects in the z=0.1-0.2 range, we needed to be more
selective in our surveys.  The third campaign was a ``snapshot
survey'' \citep{D'Vali99} in which many BL Lacs which were considered
likely candidates for TeV emission based on the same criteria as the
second survey were observed. When in a flaring state, a 10 minute
observation of Mrk\,421 or Mrk\,501 was enough to achieve a
significant detection. The selected snapshot survey targets were
therefore observed for 10 minutes each on a regular basis in the hope
of catching one of them in such a flaring state. The objects were
grouped based upon their proximity to each other on the celestial
sphere. Objects in each group were then observed consecutively so as
to minimize telescope slewing time. The remaining objects with known
redshift were chosen because they were EGRET sources (W\,Comae,
PKS\,0829+046, S4\,0954+65, 3C\,66A) or because they were a
superluminal source (OQ\,530). RGB\,J1725+118 (4U\,1722+11) was
observed because one measurement \citep{Veron93} derived a redshift of
$z = 0.018$, which would make it the closest known BL Lac object. The
estimate was however, derived from one absorption line and many papers
list its redshift as unknown (e.g., \citealp{Padovani95b}).

\section{Analysis Methods}
\label{observe}

\subsection{Telescope Configurations}
\label{tel-sec}

The VHE observations reported in this paper were made with the
atmospheric Cherenkov imaging technique \citep{Cawley95,
Reynolds93} using the 10-m optical reflector located at the Whipple
Observatory on Mt. Hopkins in Arizona (elevation 2.3 km)
\citep{Cawley90}. A camera, consisting of an array of photomultiplier 
tubes (PMTs) mounted in the focal plane of the reflector, records
images of atmospheric Cherenkov radiation from air showers produced by
gamma rays and cosmic rays. The observations reported here span five
years and the camera of the Whipple gamma-ray telescope changed
several times during that period. Table~\ref{camera-tab} outlines the
configurations of the camera. Light concentrators are reflective cones
that are mounted in front of the PMTs to improve light collection
efficiency and reduce albedo. During 1999, an intelligent trigger
``the Pattern Selection Trigger'' (PST) \citep{Bradbury99} was
installed that required three adjacent tubes to record a signal above
a certain level within a preset window. Prior to this, a trigger was
declared if any two PMTs in the camera recorded a signal above a
certain level within a preset window. The PST reduces the number of
triggers caused by fluctuations of the night-sky background and thus
allows the telescope to operate at lower energies. The mirror
reflectivity and trigger settings also changed over time. Each
observing season runs approximately from September through
June. Observations are not usually carried out in July and August
because the monsoon season, during which lightning storms strike
frequently, occurs at this time.

\subsection{Gamma-ray Selection}
\label{cuts-sec}

We characterize each Cherenkov image using a moment analysis
\citep{Reynolds93}. The roughly elliptical shape of the image is
described by the {\it length} and the {\it width} parameters, and its
location and orientation within the telescope field of view are given
by the {\it distance} and $\alpha$ parameters, respectively. We also
determine the two highest signals recorded by the PMTs ({\it max1},
{\it max2}) and the amount of light in the image {\it size}. In
addition, we can apply a cut on the third moment parameter {\it
asymmetry} to select gamma-ray candidates since the narrower tail of
the image should point back toward the source location within the
FOV. The $\alpha$ parameter tests whether the major axis of the image
is aligned with the putative source location; it does not eliminate
events whose major axes are parallel with the source location but
whose image points away from it. The {\it asymmetry} parameter is not
an efficient cut for cameras with small fields of view because the
images are often truncated.

Because of the changes in the camera discussed above in
Section~\ref{tel-sec}, the optimum cuts for selecting gamma rays
change with time. The cuts for different camera configurations are
listed in Table~\ref{cuts-tab}. They result in different
sensitivities, energy ranges, and effective areas for each
camera. This limits our ability to combine data from different
observing periods into single upper limits. Given the variable nature
of BL Lac objects however, deriving single upper limits for several
years of observation is of dubious benefit. Instead we quote upper
limits for each observing period.

\subsection{Tracking Analysis}
\label{trk-sec}

The observations presented here were all taken in the Tracking data
collection mode wherein only the on-source position is tracked, in
runs of 28 minute duration. To estimate the expected background, we
use those events that pass all of the gamma-ray selection criteria
except orientation (characterized by the $\alpha$ parameter). We use
events with values of $\alpha$ between 20$^\circ$ and 65$^\circ$ as
the background region and convert the counts to an estimated
background within the on-source region ($\alpha < 10^\circ$ or
$15^\circ$; see Table~\ref{cuts-tab}) by multiplying the number of
counts by a ratio determined from observations of non-source regions
taken at other times during the observing season. This method has been
described in detail by \citet{Catanese98}. The value of the factor
that converts the off-source counts to an on-source background
estimate varies with season due to changes in the camera sensitivity
and field of view. The estimated values for each of the observing
periods are listed in Table~\ref{analysis-tab}.

In the case of tracking analysis, to establish the significance (S) of
an excess or of a deficit, we use simple error propagation:

\begin{equation}
S = {{N_{\rm on} - r*N_{\rm bkd}}\over{
\sqrt{N_{\rm on} + r^2 * N_{\rm bkd} + N_{\rm bkd}^2 * (\Delta r)^2}}}
\end{equation}
where $N_{\rm on}$ is the number of events in the on-source region
(designated by the $\alpha$ cut in Table~\ref{cuts-tab}), $N_{\rm
bkd}$ is the number of events in the background region ($\alpha =
20^\circ - 65^\circ$), and $r \pm \Delta r$ is the tracking ratio and
its statistical uncertainty.

\subsection{Flux Upper Limit Estimation}
\label{fluX-sec}

After we select gamma-ray candidate events, we determine the
significance of any excess or deficit in the observations. If the
excess or the deficit is not statistically significant, as is the case
for all observations reported here, we calculate a 99.9\% confidence
level (C.L.) upper limit on the count rate by using the method of
\citet{Helene83}. To convert these flux upper limits to absolute 
fluxes, we first express them as a fraction of the Crab Nebula count
rate by using observations from the same observing period. Although
this method assumes a Crab-like spectrum for the BL Lacs, it corrects
for season to season variations in factors like PMT gain and mirror
reflectivity which affect the telescope response, and therefore its
gamma-ray count rate. The count rates observed for the Crab Nebula for
the observing periods reported here are given in
Table~\ref{analysis-tab}. Analysis of the Crab Nebula data shows that
for runs taken under good weather conditions, the gamma-ray count rate
does not change significantly within a season
\citep{Quinn98}. We can therefore assume that the gamma-ray count rate 
for a source can be reliably expressed in terms of the Crab Nebula
flux over the course of the season.

Once we have the flux limit expressed as a fraction of the Crab Nebula
count rate, we multiply it by the integral Crab Nebula flux (in units
of photons cm$^{-2}$ s$^{-1}$) above the peak response energy of the
observations (E$_p$). We define E$_p$ as the energy at which the
collection area folded with an E$^{-2.5}$ spectrum, that of the Crab
Nebula \citep{Hillas98}, reaches a maximum. The integral fluxes from
the Crab Nebula above E$_p$ for the different observations periods
reported here are given in Table~\ref{analysis-tab}. Upper limits are
an estimate of the flux that could be present in the data set but not
produce a significant excess. This is most accurately derived from the
count rate because that is what determines the statistical
significance of the excess. The Crab Nebula count rate and flux
uncertainties affect only the normalization, so the flux upper limits
quoted in terms of photons cm$^{-2}$ s$^{-1}$ have an uncertainty of
$\sim$25\%, mainly from the uncertainty in the Crab Nebula photon
flux.

\section{Results and Discussion}
\label{res}

\begin{figure}
\epsscale{1}
\plotone{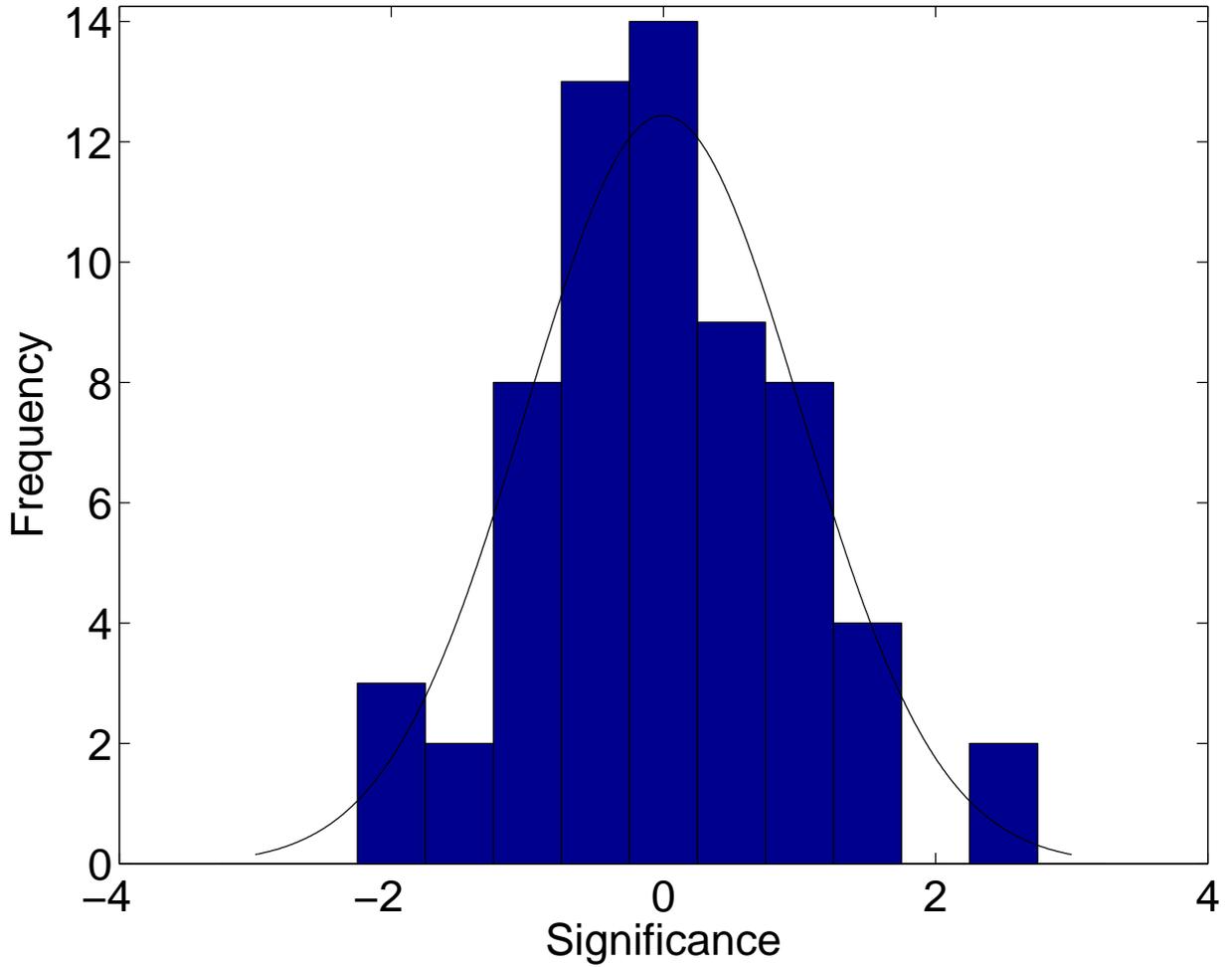}
\caption{\label{sigmas}The significance of the deficit or excess in
the detected count rate from each of the 29 BL Lac objects for each
season during which they were observed. This distribution has a mean
of 0.005 and standard deviation of 0.976. The black curve shows the
expected shape if the significances were normally distributed. This
curve fits the data at the 95\% confidence level.}
\end{figure}

Table~\ref{results-tab} summarizes the results of the observations of
the BL Lacs observed but not detected between January 1995 and July
2000 while Table~\ref{detected} summarizes the results of the
observations taken on H1426+428, 1ES1959+650 and 1ES2344+514, the
three BL Lacs that were detected during this survey. For each target
object, we list the observation exposure during each season, the
significance of the excess or deficit of these observations, the
maximum significances for a night or month of observations and the
flux upper limits expressed as fractions of the Crab count rate and in
integral flux units (assuming a Crab-like spectrum). Many of the
objects were observed over a number of different observing seasons
resulting in a range of upper limits above different values of
E$_p$. No evidence for a statistically significant excess or deficit
was seen in the detected count rate from any of the objects for any of
the time periods examined. The distribution of the significances for
each object for each observing season is shown in
Figure~\ref{sigmas}. This distribution has a mean of 0.005 with a
standard deviation of 0.976. The black curve shows the expected shape
if the significances were drawn from a Gaussian distribution. The
Kolmogorov-Smirnov test returns a 95\% probability that the data are
normally distributed.

Table~\ref{detected2} summarizes the detections of H1426+428,
1ES1959+650 and 1ES2344+514, the three target BL Lacs that have been
subsequently confirmed as TeV emitters.

In order to investigate how much the flux upper limits change if the
spectral index is not the same as that of the Crab Nebula, estimates
of the flux upper limits for the BL Lacs were made assuming source
spectral indices of -2.2 and -2.8. The integral flux from the Crab
Nebula above 300 GeV was used to scale the previously calculated upper
limits. This integral flux was assumed to remain constant when the
spectral index was changed. E$_p$ was not adjusted to account for the
response of the telescope to the different input spectra but this
effect should be very small. Table~\ref{newindices} lists the flux
upper limits for each BL Lac for each observing season when these
different source spectral indices were assumed.

\citet{Costamante02} have made predictions for the TeV flux
from fourteen of the BL Lacs included in this paper using two
different methods. In the first, an SSC model was used to fit the
multiwavelength data gathered on each of the objects while in the
second approach, a phenomenological description of the average SED of
the blazars was derived based upon their observed bolometric
luminosity \citep{Fossati98}. The resulting flux predictions from
each of the two methods were given above 300 GeV and above 1 TeV. In
order to compare the upper limits presented here with these
predictions, our upper limits that were not already derived above 300
GeV, were extrapolated to this energy. This was done by first
expressing the flux upper limit as a fraction of the Crab flux at that
energy, F$_{BLLac}$($>E_p$). This was then scaled to 300 GeV assuming
a Crab-like spectrum. Thus, an upper limit on the integral flux for
each BL Lac above 300 GeV for each observing season,
F$_{BLLac}$($>$300GeV), was calculated:

\begin{equation}\label{flux}
F_{BLLac}(>300GeV)=F_{BLLac}(>E_p)\,F_{Crab}(>E_p)\left({\displaystyle\frac{300}{E_p}}\right)^{-1.5}
\end{equation}

\noindent F$_{Crab}$($>$E$_p$) is the integral flux from the Crab Nebula 
above E$_p$ in units of photons cm$^{-2}$ s$^{-1}$, assuming an
integral spectral index of -1.5. F$_{BLLac}$($>E_p$) is the upper
limit on the flux from the BL Lac above E$_p$ expressed as a fraction
of the Crab Nebula integral flux at this energy.

When deriving their flux predictions, \citet{Costamante02} did not
take into account absorption of the gamma rays by the infra-red
background. The predictions above 300 GeV could therefore change by
factors on the order of 5 for objects at redshifts above 0.2. The
upper limits presented here were compared with these predictions and,
those of four BL Lacs, shown in Table~\ref{foscos}, were found, during
all seasons in which they were observed, to be lower than the
predicted fluxes according to the Fossati approach \citep{Fossati98}
adapted in \citep{Costamante02}.  Those of two more BL Lacs, also
listed in Table~\ref{foscos}, were found to be lower during some of
the observing periods. All of the upper limits calculated were higher
than the fluxes predicted using the one-zone, SSC model
\citep{Costamante02}. It should be noted however, that the upper
limits quoted here pertain only to the specific period during which
the observations were made. As demonstrated by \citet{Bottcher02},
spectral fitting of blazars is subject to very large uncertainties
when non-simultaneous multiwavelength data are used. Indeed, it is
also shown that, even with the best currently available simultaneous
optical - X-ray data, there is a very wide range in the predicted
fluxes above 40 GeV. Given an observed X-ray flux, the predicted
gamma-ray flux depends very sensitively on the model parameters and,
even for simultaneous data, can vary by large factors due to the
uncertainty in these parameters.

In the absence of dedicated simultaneous multiwavelength data, it is
difficult to use these data to constrain emission models. However,
many of the objects surveyed here are monitored by the All Sky Monitor
(ASM) on board the Rossi X-ray Timing Explorer. For many other TeV
blazars, the soft X-ray flux has been seen to rise during periods when
the gamma-ray flux was also high. For example, Mrk\,501 was observed
to have a higher than average flux in the ASM during 1997, the same
year during which the greatest TeV flaring activity detected to date
was observed \citep{Catanese97b,Quinn99}. In order to see if any of
the BL Lacs surveyed here were particularly active in the X-ray band
during these observations, the X-ray curves for the objects presented
here that are monitored by ASM were analyzed. If heightened X-ray
activity was detected in the absence of corresponding gamma-ray
activity, this could have interesting consequences for emission
models. Out of the 29 objects presented here, 25 of them are monitored
on a regular basis by the ASM. The nightly average light curves for
each of these were generated. The flux from each object was found to
be, on average, very low ($\lesssim$ 0.01 Crab units ) with no
evidence for dramatic flaring or for any sustained period of X-ray
activity.

Current and future observing campaigns at the Whipple Observatory make
use of BL Lac monitoring at X-ray wavelengths to try to predict when
an object might be in a higher flux state and thus detectable in the
VHE band. Elevated gamma-ray fluxes are often accompanied by a
corresponding increase in the X-ray flux (e.g. \citealp{Maraschi99};
\citealp{Jordan01}). Thus, by monitoring the X-ray activity from
blazars, we can identify periods of increased activity during which
the VHE flux may also be stronger. However, the relationship between
the X-ray and gamma-ray flux has been shown to be complicated with
gamma-ray flares being detected in the absence of X-ray flares
\citep{Holder03, Krawczynski03}, and vice-versa \citep{Rebillot03}.

Finally, it should be noted that, although none of the objects
presented here were found to have a statistically significant TeV flux
during these observations, three of the confirmed TeV blazars were
detected during this survey. When 1ES2344+514, H1426+428 and
1ES1959+650 were initially observed at Whipple, it was as part of this
BL Lac campaign and, like the objects listed here, they were not
detected. In subsequent years however, continued monitoring with
deeper exposures revealed these objects to be TeV emitters when in
more active states, although not always detectable when in their
quiescent state. Continued VHE observations of the BL Lacs presented
here, in particular those shown to have extreme properties
\citep{delaCalle03}, accompanied by monitoring of their X-ray flux
level, may reveal many more of them to be TeV emitters. Indeed, since
extreme BL Lacs, the best blazar candidates for TeV emission, have
lower luminosity at all wavelengths than their lower energy peaked
counterparts, their flux level often lies below the detection
threshold of the current generation of imaging atmospheric Cherenkov
telescopes when they are in quiescent state. As the next generation of
ground-based gamma-ray telescopes comes online with their increased
flux sensitivity, they will offer a unique opportunity to study this
low luminosity class of blazar and should detect many more of these
objects.

Future, X-ray all-sky monitor experiments like LOBSTER
\citep{Black03} and EXIST \citep{Grindlay02}, with their improved flux
sensitivity and increased bandwidth, will allow for more detailed
monitoring of the X-ray emission from blazars thus providing valuable
information which can be used to trigger observations at gamma-ray
energies. These X-ray missions, coupled with the next generation of
higher sensitivity VHE observatories such as VERITAS, HESS, MAGIC and
CANGAROO, should allow both lower power extreme BL Lacs and lower
frequency peaked blazars (LBLs and FSRQs) to be detected.

\acknowledgments 

We acknowledge the technical assistance of E. Roache and J. Melnick.
We also thank the anonymous referee for his/her comments which enabled
us to improve this paper. This research is supported by grants from
the U. S. Department of Energy, the National Science Foundation, by
Enterprise Ireland and by PPARC in the UK. This research has made use
of the NASA/IPAC Extragalactic Database (NED) which is operated by the
Jet Propulsion Laboratory, California Institute of Technology, under
contract with the National Aeronautics and Space Administration. This
research made use of the quick-look results provided by the ASM/RXTE
team.

\newpage

\begin{deluxetable}{lccll}
\tablewidth{0pt}
\tablecaption{Observed BL Lac Objects \label{object_sum}}
\tablehead{ & \colhead{R.A.} & \colhead{Dec.} \\
\colhead{Name}                                 & \colhead{(J2000)} 
                                                       & \colhead{(J2000)} 
                                                                   & \colhead{$z$} 
                                                                           & \colhead{Class\tablenotemark{a}}}
\startdata
1ES\,0033+595                                  & 00 35 52.6 & +59 50 05 & 0.086 & HBL \\
1ES\,0145+138                                  & 01 48 29.7 & +14 02 18 & 0.125 & HBL \\
RGB\,J0214+517                                 & 02 14 17.9 & +51 44 52 & 0.049 & HBL \\
3C\,66A, 1ES\,0219+428\tablenotemark{c,d}      & 02 22 39.6 & +43 02 08 & 0.444 & LBL \\
1ES\,0229+200                                  & 02 32 48.4 & +20 17 16 & 0.140 & HBL \\
1H\,0323+022, 1ES\,0323+022                    & 03 26 14.0 & +02 25 15 & 0.147 & HBL \\
EXO\,0706.1+5913, RGB\,J0710+591               & 07 10 30.0 & +59 08 20 & 0.125 & HBL \\
1ES\,0806+524                                  & 08 09 49.1 & +52 18 59 & 0.138 & HBL \\
PKS\,0829+046, RGB\,J0831+044\tablenotemark{d} & 08 31 48.9 & +04 29 39 & 0.180 & LBL \\
1ES\,0927+500                                  & 09 30 37.6 & +49 50 26 & 0.188 & HBL \\
S4\,0954+65, RGB\,J0958+655\tablenotemark{d}   & 09 58 47.2 & +65 33 55 & 0.368 & LBL \\
1ES\,1028+511                                  & 10 31 18.4 & +50 53 36 & 0.361 & HBL \\
1ES\,1118+424                                  & 11 20 48.0 & +42 12 12 & 0.124 & HBL \\
Markarian\,40                                  & 11 25 36.2 & +54 22 57 & 0.021 & HBL \\
Markarian\,180, 1ES\,1133+704                  & 11 36 26.4 & +70 09 27 & 0.045 & HBL \\
1ES\,1212+078                                  & 12 15 10.9 & +07 32 04 & 0.130 & HBL \\
ON\,325, 1ES\,1215+303                         & 12 17 52.1 & +30 07 01 & 0.130 & LBL \\
1H\,1219+301, 1ES\,1218+304                    & 12 21 21.9 & +30 10 37 & 0.182 & HBL \\
W\,Comae, 1ES\,1218+285\tablenotemark{d}       & 12 21 31.7 & +28 13 59 & 0.102 & LBL \\
MS\,1229.2+6430, RGB\,J1231+642                & 12 31 31.4 & +64 14 18 & 0.170 & HBL \\
1ES\,1239+069                                  & 12 41 48.3 & +06 36 01 & 0.150 & HBL \\
1ES\,1255+244                                  & 12 57 31.9 & +24 12 40 & 0.141 & HBL \\
OQ\,530, RGB\,J1419+543                        & 14 19 46.6 & +54 23 15 & 0.151 & LBL \\
4U\,1722+11, RGB\,J1725+118\tablenotemark{b}   & 17 25 04.3 & +11 52 15 & 0.018 & HBL \\
I\,Zw\,187, 1ES\,1727+502                      & 17 28 18.6 & +50 13 10 & 0.055 & HBL \\
1ES\,1741+196                                  & 17 43 57.8 & +19 35 09 & 0.084 & HBL \\
3C\,371, 1ES\,1807+698                         & 18 06 50.6 & +69 49 28 & 0.051 & LBL \\
BL\,Lacertae, 1ES\,2200+420\tablenotemark{d}   & 22 02 43.3 & +42 16 40 & 0.069 & LBL \\
1ES\,2321+419                                  & 23 23 52.1 & +42 10 59 & 0.059 & HBL \\
\enddata
\tablenotetext{a}{HBL = high-frequency-peaked BL Lac object; LBL = low-frequency-peaked BL Lac object.}
\tablenotetext{b}{This object is sometimes quoted as having a redshift of 0.018.  However, this is based on one absorption line \citep{Veron93} and is more commonly listed as having an unknown redshift.}
\tablenotetext{c}{Unconfirmed source of TeV gamma rays.}
\tablenotetext{d}{EGRET source of $>$100 MeV gamma rays.}
\end{deluxetable}

\begin{deluxetable}{lccccc}
\tablewidth{0pt}
\tablecaption{Whipple Camera Configurations \label{camera-tab}}
     \tablehead{    & \colhead{1995/01-} & \colhead{1997/01-} & \colhead{1997/09-} & \colhead{1998/12-} & \colhead{1999/10-} \\
\colhead{Period}    & \colhead{1996/12}  & \colhead{1997/06}  & \colhead{1998/12}  & \colhead{1999/03}  & \colhead{2000/07} }
\startdata
Number of PMTs      & 109                & 151                & 331                & 331                & 379\tablenotemark{a} \\
PMT spacing         & 0$\fdg$259         & 0$\fdg$259         & 0$\fdg$24          & 0$\fdg$24          & 0$\fdg$12 \\
Field of View       & 3$^\circ$          & 3$\fdg$3           & 4$\fdg$8           & 4$\fdg$8           & 2$\fdg$6\\
Light concentrators & yes                & yes                & no                 & yes                & yes \\
Pattern trigger     & no                 & no                 & no                 & no                 & yes \\
\enddata

\tablenotetext{a}{The camera consists of 379 inner tubes of FOV
0$\fdg$12 diameter surrounded by three circular rings of PMTs (111 in
all) of FOV 0$\fdg$26 diameter. The outer rings of tubes were not used
in this analysis and so, the parameters presented here pertain only to
the inner 379 tubes.}
\end{deluxetable}

\begin{deluxetable}{lccccc}
\tablewidth{0pt}
\tablecaption{Analysis Cuts \label{cuts-tab}}
     \tablehead{            & \colhead{1995/01-}            & \colhead{1997/01-}            &  \colhead{1997/09-}           & \colhead{1998/12-}            & \colhead{1999/10-} \\
\colhead{Period}            & \colhead{1996/12}             & \colhead{1997/06}             &  \colhead{1998/12}            & \colhead{1999/03}             & \colhead{2000/07} }
\startdata
{\it max1}\tablenotemark{a} & $>$100                        & $>$95                         & $>$60                         & $>$60                         & $>$30 \\
{\it max2}\tablenotemark{a} & $>$80                         & $>$45                         & $>$40                         & $>$40                         & $>$30 \\
{\it size}\tablenotemark{a} & $>$400                        & N.A.\tablenotemark{b}         & N.A.                          & N.A.                          &   N.A. \\
{\it length}                & $> 0 \fdg 16$                 & $> 0 \fdg 16$                 & $> 0 \fdg 16$                 & $> 0 \fdg 16$                 & $> 0 \fdg 13$                \\
                            &                $< 0 \fdg 30$  &                $< 0 \fdg 33$  &                $< 0 \fdg 50$  &                $< 0 \fdg 50$  &                $< 0 \fdg 25$ \\
{\it width}                 & $> 0 \fdg 073$                & $> 0 \fdg 073$                & $> 0 \fdg 073$                & $> 0 \fdg 073$                & $> 0 \fdg 05 $                \\
                            &                 $< 0 \fdg 15$ &                 $< 0 \fdg 16$ &                 $< 0 \fdg 16$ &                 $< 0 \fdg 16$ &                 $< 0 \fdg 12$ \\
{\it distance}              & $> 0 \fdg 51$                 & $> 0 \fdg 51$                 & $> 0 \fdg 51$                 & $> 0 \fdg 51$                 & $> 0 \fdg 40$                \\
                            &                $< 1 \fdg 10$  &                $< 1 \fdg 17$  &                $< 1 \fdg 55$  &                $< 1 \fdg 55$  &                $< 1 \fdg 00$ \\
{\it alpha}                 & $< 15^\circ$                  & $< 15^\circ$                  & $< 15^\circ$                  & $< 15^\circ$                  &   $< 15^\circ$ \\
{\it asymmetry}             & N.A.                          & $> 0^\circ$                   & $> 0^\circ$                   & $> 0^\circ$                   & $> 0^\circ$ \\
{\it length over size}      & N.A.                          & N.A.                          & N.A.                          & N.A.                          &   $< 0.0004$ \\
\enddata
\tablenotetext{a}{Quantities are in units of digital counts (d.c.):  1 d.c. $\approx$ 1 photoelectron.}
\tablenotetext{b}{N.A. means the cut was not applied.}
\end{deluxetable}

\begin{deluxetable}{rcccc}
\tablewidth{0pt}
\tablecaption{Analysis Parameters \label{analysis-tab}}
\tablehead{       & \colhead{Tracking}& \colhead{Crab rate}      & \colhead{Peak Response}& \colhead{Integral} \\
\colhead{Period}  & \colhead{Ratio}   & \colhead{($\gamma$/min)} & \colhead{Energy}       & \colhead{Crab flux\tablenotemark{a}} }
\startdata
1995/01 - 1995/08 & 0.292 $\pm$ 0.005 & 2.08 $\pm$ 0.15          & 300 GeV                & 1.26 \\
1995/10 - 1996/07 & 0.292 $\pm$ 0.004 & 1.58 $\pm$ 0.05          & 350 GeV                & 1.05 \\
1996/10 - 1996/12 & 0.316 $\pm$ 0.004 & 1.69 $\pm$ 0.07          & 350 GeV                & 1.05 \\
1997/01 - 1997/06 & 0.345 $\pm$ 0.005 & 2.30 $\pm$ 0.10          & 350 GeV                & 1.05 \\
1998/01 - 1998/12 & 0.366 $\pm$ 0.002 & 1.94 $\pm$ 0.15          & 500 GeV                & 0.60 \\
1998/12 - 1999/03 & 0.367 $\pm$ 0.004 & 2.62 $\pm$ 0.26          & 400 GeV                & 0.84 \\
1999/10 - 2000/07 & 0.312 $\pm$ 0.002 & 2.64 $\pm$ 0.12          & 430 GeV                & 0.76 \\
\enddata
\tablenotetext{a}{Fluxes are quoted in units of 10$^{-10}$ photons cm$^{-2}$ s$^{-1}$ above the corresponding peak response energy.}
\end{deluxetable}

\begin{deluxetable}{lrrrrrrrr}
\tablewidth{0pt}

\tablecaption{The observation results for each object, for 
each observing period during which it was observed. For each BL Lac,
the total combined significance for each period is given along with
the maximum statistical significance seen over any one night and any
month during that period. The flux upper limits are also presented for
each observing period both in absolute terms (Flux units, f.u.,
10$^{-11}$ cm$^{-2}$ s$^{-1}$) and in Crab units
(c.u.). \label{results-tab}}

\tablehead{    &\colhead{Observation}& \colhead{Exp.}  &                    & \colhead{Max.\ $\sigma$} & \colhead{Max.\ $\sigma$} & \colhead{Flux}   & \colhead{Flux}\\
\colhead{Object} & \colhead{Period}  & \colhead{(hrs)} & \colhead{$\sigma$} & \colhead{Month}          & \colhead{Night}          & \colhead{(c.u.)} & \colhead{(f.u.)} }
\startdata
1ES\,0033+595    & 1995/12           &  1.85           & -0.59              & -0.59  &  0.19        &  $<$0.200        & $<$2.10 \\
1ES\,0145+138    & 1996/10 - 1996/11 &  7.85           & -1.01              &  0.08  &  1.77        &  $<$0.093        & $<$0.98 \\
                 & 1998/11 - 1998/12 &  2.29           &  0.22              &  0.63  &  0.63        &  $<$0.512        & $<$3.50 \\
                 & 1998/12 - 1999/01 &  1.98           & -0.50              &  0.38  &  1.31        &  $<$0.357        & $<$3.34 \\
RGB\,J0214+517   & 1999/12 - 2000/01 &  6.01           &  0.29              &  0.36  &  1.58        &  $<$0.165        & $<$1.45 \\
3C\,66A          & 1995/10 - 1995/11 &  8.00           & -2.00              & -1.18  &  0.82        &  $<$0.056        & $<$0.59 \\
1ES\,0229+200    & 1996/11 - 1996/12 &  7.85           &  0.15              &  0.57  &  1.37        &  $<$0.113        & $<$1.19 \\
                 & 1998/11 - 1998/12 &  2.30           & -1.08              & -0.78  &  0.48        &  $<$0.326        & $<$2.23 \\
                 & 1998/12 - 1999/01 &  1.78           & -0.40              &  0.66  &  0.74        &  $<$0.403        & $<$3.76 \\
1H\,0323+022     & 1996/11 - 1996/12 & 10.18           &  1.02              &  1.02  &  1.96        &  $<$0.181        & $<$1.90 \\
                 & 1997/01           &  0.91           &  0.20              &  0.20  &  0.20        &  $<$0.298        & $<$3.13 \\
                 & 1998/12 - 1999/01 &  3.18           &  1.69              &  1.73  &  1.85        &  $<$0.509        & $<$4.75 \\
EXO\,0706.1+5913 & 1996/12           &  5.55           & -1.16              & -1.16  &  0.17        &  $<$0.087        & $<$0.91 \\
                 & 1997/01 - 1997/03 &  3.69           &  0.76              &  0.79  &  1.46        &  $<$0.161        & $<$1.69 \\
                 & 1998/11           &  1.83           & -0.40              & -0.40  &  1.49        &  $<$0.524        & $<$3.58 \\
                 & 1998/12 - 1999/02 &  1.90           &  0.07              &  1.36  &  1.71        &  $<$0.459        & $<$4.29 \\
1ES\,0806+524    & 1996/02 - 1996/03 &  5.57           &  0.46              &  0.52  &  1.04        &  $<$0.104        & $<$1.09 \\
                 & 2000/01 - 2000/03 &  4.16           & -0.29              &  0.69  &  0.69        &  $<$1.293        & $<$11.4 \\
PKS\,0829+046    & 1995/01 - 1995/04 & 11.07           &  1.25              &  1.73  &  3.14        &  $<$0.117        & $<$1.47 \\
1ES\,0927+500    & 1996/12           &  5.08           & -1.92              & -1.92  &  0.39        &  $<$0.064        & $<$0.67 \\
                 & 1997/01 - 1997/04 &  5.04           & -1.03              &  0.22  &  1.23        &  $<$0.076        & $<$0.80 \\
S4\,0954+65      & 1995/02 - 1995/03 &  3.70           & -1.09              & -0.50  & -0.11        &  $<$0.096        & $<$1.21 \\
1ES\,1028+511    & 1998/12 - 1999/02 &  4.43           &  0.57              &  1.36  &  2.03        &  $<$0.287        & $<$2.68 \\
1ES\,1118+424    & 1998/02 - 1998/04 &  7.30           & -0.25              &  1.29  &  1.80        &  $<$0.218        & $<$1.49 \\
                 & 1998/12 - 1999/02 &  3.60           &  0.27              &  1.04  &  1.81        &  $<$0.310        & $<$2.90 \\
                 & 2000/01 - 2000/05 &  6.97           & -0.62              &  1.29  &  1.61        &  $<$0.116        & $<$1.02 \\
Markarian\,40    & 2000/01 - 2000/04 & 10.16           &  2.59              &  1.66  &  1.56        &  $<$0.206        & $<$1.81 \\
Markarian\,180   & 1995/01 - 1995/04 &  5.55           & -0.10              &  0.45  &  1.07        &  $<$0.108        & $<$1.36 \\
                 & 1995/12 - 1996/05 & 20.46           & -0.26              &  1.01  &  1.70        &  $<$0.105        & $<$1.10 \\
                 & 1997/01           &  0.79           & -0.17              & -0.17  & -0.17        &  $<$0.303        & $<$3.18 \\
1ES\,1212+078    & 1999/02           &  1.13           &  0.44              &  0.44  &  1.83        &  $<$0.778        & $<$7.26 \\
                 & 2000/01 - 2000/05 &  3.70           &  1.30              &  1.52  &  1.73        &  $<$0.321        & $<$2.82 \\
ON\,325          & 1999/02           &  0.97           &  1.27              &  1.27  &  1.20        &  $<$0.882        & $<$8.23 \\
                 & 2000/01 - 2000/05 &  5.05           &  0.88              &  1.94  &  1.62        &  $<$0.215        & $<$1.89 \\
1H\,1219+301     & 1995/01 - 1995/05 &  2.77           &  2.71              &  2.90  &  2.95        &  $<$0.226        & $<$2.85 \\
                 & 1997/02 - 1997/06 & 11.27           &  0.99              &  1.55  &  1.97        &  $<$0.079        & $<$0.83 \\
                 & 1998/01 - 1998/03 &  1.38           & -1.96              & -1.27  & -0.32        &  $<$0.356        & $<$2.43 \\
                 & 1998/12 - 1999/02 &  2.94           & -0.08              &  0.48  &  0.88        &  $<$0.296        & $<$2.77 \\
                 & 2000/01 - 2000/04 &  3.69           &  0.04              &  1.48  &  1.18        &  $<$0.191        & $<$1.68 \\
W\,Comae         & 1995/02 - 1995/04 & 14.33           & -0.57              & -0.11  &  1.14        &  $<$0.052        & $<$0.66 \\
                 & 1996/01 - 1996/05 & 15.73           & -0.29              &  0.24  &  1.38        &  $<$0.055        & $<$0.58 \\
                 & 1999/01 - 1999/02 &  4.43           & -0.03              &  0.58  &  1.91        &  $<$0.312        & $<$2.92 \\
                 & 2000/01 - 2000/04 &  4.72           & -0.58              &  1.01  &  1.76        &  $<$0.148        & $<$1.30 \\
MS\,1229.2+6430  & 1995/02 - 1995/04 &  1.39           &  1.32              &  1.08  &  1.08        &  $<$0.286        & $<$3.60 \\
                 & 1999/02           &  2.04           & -0.76              & -0.76  &  1.18        &  $<$0.446        & $<$4.16 \\
                 & 2000/01 - 2000/05 &  6.01           &  0.35              &  1.72  &  1.72        &  $<$0.170        & $<$1.50 \\
1ES\,1239+069    & 1999/01 - 1999/02 &  1.73           &  0.78              &  0.69  &  1.74        &  $<$0.616        & $<$6.04 \\
                 & 2000/01 - 2000/05 &  5.08           &  0.11              &  1.19  &  1.42        &  $<$0.197        & $<$1.73 \\
1ES\,1255+244    & 1997/02 - 1997/05 &  5.54           &  1.19              &  1.01  &  1.35        &  $<$0.112        & $<$1.18 \\
                 & 1998/03           &  0.46           &  0.13              &  0.13  &  0.13        &  $<$1.112        & $<$7.60 \\
                 & 1999/02           &  1.73           &  0.15              &  0.15  &  1.10        &  $<$0.508        & $<$4.75 \\
                 & 2000/01 - 2000/05 &  4.16           & -0.54              &  1.30  &  1.41        &  $<$0.164        & $<$1.45 \\
OQ\,530          & 1995/03 - 1995/05 &  7.39           & -0.73              & -0.15  &  0.76        &  $<$0.058        & $<$0.73 \\
4U\,1722+11      & 1995/04 - 1995/05 &  2.77           & -0.08              &  0.28  &  0.70        &  $<$0.124        & $<$1.56 \\
I\,Zw\,187       & 1995/03 - 1995/04 &  2.31           & -1.27              & -0.07  &  0.63        &  $<$0.086        & $<$1.08 \\
                 & 1996/04 - 1996/05 &  2.32           &  0.61              &  0.85  &  1.19        &  $<$0.150        & $<$1.58 \\
1ES\,1741+196    & 1996/05 - 1996/07 &  9.23           & -1.02              &  0.46  &  2.07        &  $<$0.053        & $<$0.56 \\
                 & 1998/05           &  0.46           & -0.08              & -0.08  & -0.08        &  $<$1.168        & $<$7.99 \\
3C\,371          & 1995/05 - 1995/06 & 13.04           &  0.41              &  0.41  &  1.68        &  $<$0.190        & $<$1.23 \\
BL\,Lacertae     & 1995/07           &  4.62           &  1.07              &  1.09  &  1.09        &  $<$0.109        & $<$1.37 \\
                 & 1995/10 - 1995/11 & 39.09           & -1.48              & -0.21  &  0.85        &  $<$0.038        & $<$0.40 \\
                 & 1998/05 - 1998/06 &  0.92           &  0.47              &  0.71  &  0.71        &  $<$1.722        & $<$8.02 \\
1ES\,2321+419    & 1995/10 - 1995/11 &  6.42           & -1.07              &  1.50  &  1.50        &  $<$0.101        & $<$1.06 \\
\enddata
\end{deluxetable}

\begin{deluxetable}{lrrrrrrr}
\tablewidth{0pt}

\tablecaption{The observation results for the three detected BL Lacs
that were originally observed as part of this survey. For each
observing period during which they were observed, prior to detection,
the total combined significance is given along with the maximum
statistical significance seen over any one night and any month during
that period. The flux upper limits are also presented for each
observing period in absolute terms (Flux units, f.u., 10$^{-11}$
cm$^{-2}$ s$^{-1}$). \label{detected}}

\tablehead{    &\colhead{Observation}& \colhead{Exp.}  &                    & \colhead{Max.\ $\sigma$} & \colhead{Max.\ $\sigma$} & \colhead{Flux}\\
\colhead{Object} & \colhead{Period}  & \colhead{(hrs)} & \colhead{$\sigma$} & \colhead{Month}          & \colhead{Night}          & \colhead{(f.u.)} }
\startdata
H\,1426+428      & 1995/06 - 1995/07 &  3.48           &  2.11              &  2.11  &  2.05        &  $<$2.2  \\
                 & 1997/02 - 1997/06 & 13.16           &  1.70              &  2.18  &  1.62        &  $<$0.8  \\
                 & 1998/04           &  0.87           &  1.70              &  1.70  &  1.99        &  $<$6.7  \\
1ES\,1959+650    & 1995/06           &  7.21           &  1.02              &  0.93  &  1.17        &  $<$1.4  \\
                 & 1996/05 - 1996/07 &  3.25           &  0.33              &  0.68  &  1.10        &  $<$1.5  \\
                 & 1998/07           &  0.16           & -0.11              & -0.11  & -0.11        &  $<$12.6 \\
1ES\,2344+514    & 1995/10 - 1996/01 & 20.50           &  5.82              &  6.50  &  5.80        &  ---\tablenotemark{a}\\
\enddata
\tablenotetext{a}{1ES2344+514 was detected during the first observing
season in which it was observed. See Table~\ref{detected2} for the
details.}
\end{deluxetable}

\begin{deluxetable}{lrcrrr}
\tablewidth{0pt}

\tablecaption{The flux upper limits for each object, for each
observing period during which it was observed when input spectra with
different spectral indices ($\alpha$) were assumed. The flux upper
limits are given in absolute terms (Flux units, f.u., 10$^{-11}$
cm$^{-2}$ s$^{-1}$). \label{newindices}}

\tablehead{      &                      & \colhead{Peak}     &\colhead{Flux}           &\colhead{Flux}           &\colhead{Flux}         \\
                 & \colhead{Observation}& \colhead{Response} &\colhead{$\alpha: -2.2$} &\colhead{$\alpha: -2.5$} &\colhead{$\alpha: -2.8$}\\
\colhead{Object} & \colhead{Period}     & \colhead{Energy}   &\colhead{(f.u.)}         &\colhead{(f.u.)}         &\colhead{(f.u.)}         }
\startdata
1ES\,0033+595    & 1995/12           & 350  & $<$2.1476 &  $<$2.1000 &  $<$1.9579 \\
1ES\,0145+138    & 1996/10 - 1996/11 & 350  & $<$1.0022 &  $<$0.9800 &  $<$0.9137 \\
                 & 1998/11 - 1998/12 & 500  & $<$4.0605 &  $<$3.5000 &  $<$2.9886 \\
                 & 1998/12 - 1999/01 & 400  & $<$3.6323 &  $<$3.3400 &  $<$3.0565 \\
RGB\,J0214+517   & 1999/12 - 2000/01 & 430  & $<$1.6103 &  $<$1.4500 &  $<$1.2974 \\
3C\,66A          & 1995/10 - 1995/11 & 350  & $<$0.6034 &  $<$0.5900 &  $<$0.5501 \\
1ES\,0229+200    & 1996/11 - 1996/12 & 350  & $<$1.2170 &  $<$1.1900 &  $<$1.1095 \\
                 & 1998/11 - 1998/12 & 500  & $<$2.5871 &  $<$2.2300 &  $<$1.9042 \\
                 & 1998/12 - 1999/01 & 400  & $<$4.0891 &  $<$3.7600 &  $<$3.4408 \\
1H\,0323+022     & 1996/11 - 1996/12 & 350  & $<$1.9431 &  $<$1.9000 &  $<$1.7714 \\
                 & 1997/01           & 350  & $<$3.2010 &  $<$3.1300 &  $<$2.9182 \\
                 & 1998/12 - 1999/01 & 400  & $<$5.1657 &  $<$4.7500 &  $<$4.3468 \\
EXO\,0706.1+5913 & 1996/12           & 350  & $<$0.9306 &  $<$0.9100 &  $<$0.8484 \\
                 & 1997/01 - 1997/03 & 350  & $<$1.7283 &  $<$1.6900 &  $<$1.5756 \\
                 & 1998/11           & 500  & $<$4.1533 &  $<$3.5800 &  $<$3.0569 \\
                 & 1998/12 - 1999/02 & 400  & $<$4.6655 &  $<$4.2900 &  $<$3.9258 \\
1ES\,0806+524    & 1996/02 - 1996/03 & 350  & $<$1.1147 &  $<$1.0900 &  $<$1.0162 \\
                 & 2000/01 - 2000/03 & 430  &$<$12.6599 & $<$11.4000 & $<$10.2005 \\
PKS\,0829+046    & 1995/01 - 1995/04 & 300  & $<$1.4700 &  $<$1.4700 &  $<$1.4700 \\
1ES\,0927+500    & 1996/12           & 350  & $<$0.6852 &  $<$0.6700 &  $<$0.6247 \\
                 & 1997/01 - 1997/04 & 350  & $<$0.8181 &  $<$0.8000 &  $<$0.7459 \\
S4\,0954+65      & 1995/02 - 1995/03 & 300  & $<$1.2100 &  $<$1.2100 &  $<$1.2100 \\
1ES\,1028+511    & 1998/12 - 1999/02 & 400  & $<$2.9146 &  $<$2.6800 &  $<$2.4525 \\
1ES\,1118+424    & 1998/02 - 1998/04 & 500  & $<$1.7286 &  $<$1.4900 &  $<$1.2723 \\
                 & 1998/12 - 1999/02 & 400  & $<$3.1538 &  $<$2.9000 &  $<$2.6538 \\
                 & 2000/01 - 2000/05 & 430  & $<$1.1327 &  $<$1.0200 &  $<$0.9127 \\
Markarian\,40    & 2000/01 - 2000/04 & 430  & $<$2.0100 &  $<$1.8100 &  $<$1.6196 \\
Markarian\,180   & 1995/01 - 1995/04 & 300  & $<$1.3600 &  $<$1.3600 &  $<$1.3600 \\
                 & 1995/12 - 1996/05 & 350  & $<$1.1249 &  $<$1.1000 &  $<$1.0256 \\
                 & 1997/01           & 350  & $<$3.2521 &  $<$3.1800 &  $<$2.9648 \\
1ES\,1212+078    & 1999/02           & 400  & $<$7.8954 &  $<$7.2600 &  $<$6.6437 \\
                 & 2000/01 - 2000/05 & 430  & $<$3.1317 &  $<$2.8200 &  $<$2.5233 \\
ON\,325          & 1999/02           & 400  & $<$8.9503 &  $<$8.2300 &  $<$7.5314 \\
                 & 2000/01 - 2000/05 & 430  & $<$2.0989 &  $<$1.8900 &  $<$1.6911 \\
1H\,1219+301     & 1995/01 - 1995/05 & 300  & $<$2.8500 &  $<$2.8500 &  $<$2.8500 \\
                 & 1997/02 - 1997/06 & 350  & $<$0.8488 &  $<$0.8300 &  $<$0.7738 \\
                 & 1998/01 - 1998/03 & 500  & $<$2.8191 &  $<$2.4300 &  $<$2.0750 \\
                 & 1998/12 - 1999/02 & 400  & $<$3.0124 &  $<$2.7700 &  $<$2.5349 \\
                 & 2000/01 - 2000/04 & 430  & $<$1.8657 &  $<$1.6800 &  $<$1.5032 \\
W\,Comae         & 1995/02 - 1995/04 & 300  & $<$0.6600 &  $<$0.6600 &  $<$0.6600 \\
                 & 1996/01 - 1996/05 & 350  & $<$0.5932 &  $<$0.5800 &  $<$0.5408 \\
                 & 1999/01 - 1999/02 & 400  & $<$3.1756 &  $<$2.9200 &  $<$2.6721 \\
                 & 2000/01 - 2000/04 & 430  & $<$1.4437 &  $<$1.3000 &  $<$1.1632 \\
MS\,1229.2+6430  & 1995/02 - 1995/04 & 300  & $<$3.6000 &  $<$3.6000 &  $<$3.6000 \\
                 & 1999/02           & 400  & $<$4.5241 &  $<$4.1600 &  $<$3.8069 \\
                 & 2000/01 - 2000/05 & 430  & $<$1.6658 &  $<$1.5000 &  $<$1.3422 \\
1ES\,1239+069    & 1999/01 - 1999/02 & 400  & $<$6.5686 &  $<$6.0400 &  $<$5.5273 \\
                 & 2000/01 - 2000/05 & 430  & $<$1.9212 &  $<$1.7300 &  $<$1.5480 \\
1ES\,1255+244    & 1997/02 - 1997/05 & 350  & $<$1.2068 &  $<$1.1800 &  $<$1.1001 \\
                 & 1998/03           & 500  & $<$8.8171 &  $<$7.6000 &  $<$6.4896 \\
                 & 1999/02           & 400  & $<$5.1657 &  $<$4.7500 &  $<$4.3468 \\
                 & 2000/01 - 2000/05 & 430  & $<$1.6103 &  $<$1.4500 &  $<$1.2974 \\
OQ\,530          & 1995/03 - 1995/05 & 300  & $<$0.7300 &  $<$0.7300 &  $<$0.7300 \\
4U\,1722+11      & 1995/04 - 1995/05 & 300  & $<$1.5600 &  $<$1.5600 &  $<$1.5600 \\
I\,Zw\,187       & 1995/03 - 1995/04 & 300  & $<$1.0800 &  $<$1.0800 &  $<$1.0800 \\
                 & 1996/04 - 1996/05 & 350  & $<$1.6158 &  $<$1.5800 &  $<$1.4731 \\
1ES\,1741+196    & 1996/05 - 1996/07 & 350  & $<$0.5727 &  $<$0.5600 &  $<$0.5221 \\
                 & 1998/05           & 500  & $<$9.2695 &  $<$7.9900 &  $<$6.8226 \\
3C\,371          & 1995/05 - 1995/06 & 300  & $<$1.2300 &  $<$1.2300 &  $<$1.2300 \\
BL\,Lacertae     & 1995/07           & 300  & $<$1.3700 &  $<$1.3700 &  $<$1.3700 \\
                 & 1995/10 - 1995/11 & 350  & $<$0.4091 &  $<$0.4000 &  $<$0.3729 \\
                 & 1998/05 - 1998/06 & 500  & $<$9.3043 &  $<$8.0200 &  $<$6.8482 \\
1ES\,2321+419    & 1995/10 - 1995/11 & 350  & $<$1.0840 &  $<$1.0600 &  $<$0.9883 \\
\enddata
\end{deluxetable}

\begin{deluxetable}{llccl}
\tablewidth{0pt}

\tablecaption{Summary of the detections of the three objects,
H1426+428, 1ES1959+650 and 1ES2344+514, that were detected during this
survey. The observing season during which they were detected, the peak
response energy for that season and, where available, the detection
integral flux in units of 10$^{-11}$ photons cm$^{-2}$ s$^{-1}$) above
the corresponding peak response energy, along with a reference to the
detection paper, are given for each object.\label{detected2}}

\tablehead{      &\colhead{Observing Season} & \colhead{Peak Response} & \colhead{Integral}    & \colhead{Detection} \\
\colhead{Object} & \colhead{of Detection}    & \colhead{Energy}        & \colhead{Flux}        & \colhead{Paper} }   
\startdata
H\,1426+428      & 2000/10 - 2001/07         &  280 GeV                &  2.0 $\pm$ 0.3        & \citet{Horan02}     \\
1ES\,1959+650    & 2001/10 - 2002/07         &  600 GeV                &  ---\tablenotemark{a} & \citet{Holder03}    \\
1ES\,2344+514    & 1995/10 - 1996/07         &  350 GeV                &  1.1 $\pm$ 0.4        & \citet{Catanese98}  \\
\enddata
\tablenotetext{a}{No flux was quoted in the detection paper due to
difficulties in performing a spectral analysis because of a decrease
in telescope efficiency during the course of the observations.}
\end{deluxetable}

\begin{deluxetable}{lrccc}
\tablewidth{0pt}

\tablecaption{The 30 Upper Limits scaled to 300 GeV compared with the
flux estimates of \citet{Costamante02} (where available). Both the
flux predictions and upper limits are given in absolute flux units,
f.u., 10$^{-11}$ cm$^{-2}$ s$^{-1}$). As was described in the
text, the Upper Limits from \citet{Costamante02} were calculated using
two different approaches, that of Costamante (Cos) and that of Fossati
(Fos).
\label{foscos}}

\tablehead{      &                  &                & \colhead{F ($>$300 GeV)}                & \colhead{F ($>$ 300 GeV)}             \\
               \colhead{Observation}& \colhead{Exp.} & \colhead{(f.u.)}                        & \colhead{(f.u.)}  \\
\colhead{Object} & \colhead{Period} & \colhead{(hrs)}& \colhead{Fos/Cos}                       & \colhead{Extrapolated}                   }
\startdata
1ES\,0033+595    &           1995/12 &  1.85         & 2.04 / 0.25                             & $<$  2.64  \\
1ES\,0145+138    & 1996/10 - 1996/11 &  7.85         &  --- / ---                              & $<$  1.23  \\
                 & 1998/11 - 1998/12 &  2.29         &                                         & $<$  6.58  \\
                 & 1998/12 - 1999/01 &  1.98         &                                         & $<$  4.60  \\
RGB\,J0214+517   & 1999/12 - 2000/01 &  6.01         & 5.93 / 0.07                             & $<$  2.14  \\
3C\,66A          & 1995/10 - 1995/11 &  8.00         & 0.14 / ---                              & $<$  0.74  \\
1ES\,0229+200    & 1996/11 - 1996/12 &  7.85         & 0.96 / 0.31                             & $<$  1.49  \\
                 & 1998/11 - 1998/12 &  2.30         &                                         & $<$  4.19  \\
                 & 1998/12 - 1999/01 &  1.78         &                                         & $<$  5.20  \\
1H\,0323+022     & 1996/11 - 1996/12 & 10.18         & 0.84 / 0.01                             & $<$  2.39  \\
                 &           1997/01 &  0.91         &                                         & $<$  3.94  \\
                 & 1998/12 - 1999/01 &  3.18         &                                         & $<$  6.56  \\
EXO\,0706.1+5913 & 1996/12           &  5.55         &  --- / ---                              & $<$  1.15  \\
                 & 1997/01 - 1997/03 &  3.69         &                                         & $<$  2.13  \\
                 & 1998/11           &  1.83         &                                         & $<$  6.73  \\
                 & 1998/12 - 1999/02 &  1.90         &                                         & $<$  5.92  \\
1ES\,0806+524    & 1996/02 - 1996/03 &  5.57         & 1.36 / ---                              & $<$  1.37  \\
                 & 2000/01 - 2000/03 &  4.16         &                                         & $<$ 16.80  \\
PKS\,0829+046    & 1995/01 - 1995/04 & 11.07         &  --- / ---                              & $<$  1.47  \\
1ES\,0927+500    & 1996/12           &  5.08         &  --- / ---                              & $<$  0.85  \\
                 & 1997/01 - 1997/04 &  5.04         &                                         & $<$  1.00  \\
S4\,0954+65      & 1995/02 - 1995/03 &  3.70         &  --- / ---                              & $<$  1.21  \\
1ES\,1028+511    & 1998/12 - 1999/02 &  4.43         & 0.43 / ---                              & $<$  3.70  \\
1ES\,1118+424    & 1998/02 - 1998/04 &  7.30         &  --- / ---                              & $<$  2.80  \\
                 & 1998/12 - 1999/02 &  3.60         &                                         & $<$  4.00  \\
                 & 2000/01 - 2000/05 &  6.97         &                                         & $<$  1.51  \\
Markarian\,40    & 2000/01 - 2000/04 & 10.16         &  --- / ---                              & $<$  2.68  \\
Markarian\,180   & 1995/01 - 1995/04 &  5.55         & 8.50 / 0.03                             & $<$  1.36  \\
                 & 1995/12 - 1996/05 & 20.46         &                                         & $<$  1.39  \\
                 &           1997/01 &  0.79         &                                         & $<$  4.00  \\
1ES\,1212+078    & 1999/02           &  1.13         &  --- / ---                              & $<$ 10.03  \\
                 & 2000/01 - 2000/05 &  3.70         &                                         & $<$  4.17  \\
ON\,325          &           1999/02 &  0.97         & 0.16 / ---                              & $<$ 11.37  \\
                 & 2000/01 - 2000/05 &  5.05         &                                         & $<$  2.79  \\
1H\,1219+301     & 1995/01 - 1995/05 &  2.77         & 0.67 / 0.16                             & $<$  2.85  \\
                 & 1997/02 - 1997/06 & 11.27         &                                         & $<$  1.04  \\
                 & 1998/01 - 1998/03 &  1.38         &                                         & $<$  4.57  \\
                 & 1998/12 - 1999/02 &  2.94         &                                         & $<$  3.82  \\
                 & 2000/01 - 2000/04 &  3.69         &                                         & $<$  2.48  \\
W\,Comae         & 1995/02 - 1995/04 & 14.33         &  --- / ---                              & $<$  0.66  \\
                 & 1996/01 - 1996/05 & 15.73         &                                         & $<$  0.73  \\
                 & 1999/01 - 1999/02 &  4.43         &                                         & $<$  4.02  \\
                 & 2000/01 - 2000/04 &  4.72         &                                         & $<$  1.92  \\
MS\,1229.2+6430  & 1995/02 - 1995/04 &  1.39         &  --- / ---                              & $<$  3.60  \\
                 & 1999/02           &  2.04         &                                         & $<$  5.75  \\
                 & 2000/01 - 2000/05 &  6.01         &                                         & $<$  2.21  \\
1ES\,1239+069    & 1999/01 - 1999/02 &  1.73         &  --- / ---                              & $<$  7.94  \\
                 & 2000/01 - 2000/05 &  5.08         &                                         & $<$  2.56  \\
1ES\,1255+244    & 1997/02 - 1997/05 &  5.54         &  --- / ---                              & $<$  1.48  \\
                 & 1998/03           &  0.46         &                                         & $<$ 14.28  \\
                 & 1999/02           &  1.73         &                                         & $<$  6.55  \\
                 & 2000/01 - 2000/05 &  4.16         &                                         & $<$  2.13  \\
OQ\,530          & 1995/03 - 1995/05 &  7.39         &  --- / ---                              & $<$  0.73  \\
4U\,1722+11      & 1995/04 - 1995/05 &  2.77         & 12.8 / 0.015                            & $<$  1.56  \\
I\,Zw\,187       & 1995/03 - 1995/04 &  2.31         & 5.19 / 0.07                             & $<$  1.08  \\
                 & 1996/04 - 1996/05 &  2.32         &                                         & $<$  1.98  \\
1ES\,1741+196    & 1996/05 - 1996/07 &  9.23         & 3.59 / 0.29                             & $<$  0.70  \\
                 &           1998/05 &  0.46         &                                         & $<$ 15.00  \\
3C\,371          & 1995/05 - 1995/06 & 13.04         &  --- / ---                              & $<$  1.23  \\
BL\,Lacertae     &           1995/07 &  4.62         & 3.32 / 0.17                             & $<$  1.37  \\
                 & 1995/10 - 1995/11 & 39.09         &                                         & $<$  0.50  \\
                 & 1998/05 - 1998/06 &  0.92         &                                         & $<$ 22.12  \\
1ES\,2321+419    & 1995/10 - 1995/11 &  6.42         &  --- / ---                              & $<$  1.33  \\
\enddata
\end{deluxetable}

\end{document}